# A Computer Vision Based Beamforming Scheme for Millimeter Wave Communication in LOS Scenarios

Tianqi Xiang*, Yaxin Wang*, Huiwen Li*, Boren Guo*, Xin Zhang*
*Wireless Theories and Technologies Lab
Bejing University of Posts and Telecommunications
Beijing 100876, China
Emails: xiangtianqi@sina.com; Godsword98@126.com; lhw5799@bupt.edu.cn; plume@bupt.edu.cn; zhangxin@bupt.edu.cn

***Abstract*—A novel location-aware beamforming scheme for millimeter wave communication is proposed for line of sight (LOS) and low mobility scenarios, in which computer vision is introduced to derive the required position or spatial angular information from the image or video captured by camera(s) co-located with mmWave antenna array at base stations. A wireless coverage model is built to investigate the coverage performance and influence of positioning accuracy achieved by convolutional neural network (CNN) for image processing. In addition, videos could be intentionally blurred, or even low-resolution videos could be directly applied, to protect users' privacy with acceptable positioning precision, lower computation complexity and lower camera cost. It is proved by simulations that the beamforming scheme is practicable and the mainstream CNN we employed is sufficient in both aspects of beam directivity accuracy and processing speed in frame per second.**

## I. Introduction

It is widely agreed that due to explosively growing demands on data rates of mobile communication, larger bandwidths are needed, and thus higher frequencies are considered to be applied to the next generation mobile communication systems, which is the reason millimeter wave becomes a research hot spot. Due to its high frequency, mmwave has a better performance in beam directivity, and however, poorer one in propagation losses. Therefore, beamforming techniques are being used to overcome the poor propagation conditions, but at the cost of entailing more stringent appeal of spatial channel information.

Location-aware communication aims to utilize the location information for enhancing the usage of spatial dimension including beamforming. 5G dense network has been proposed for learning location information, which in turn can be used efficiently in radio network including geometric beamforming[1]. Location-awareness has also been proved beneficial in Internet of Things (IoT) with multi-connectivity at 28GHz[2]. And in high speed railway scenarios, it is put forward that beams can be generated and selected in a low-complexity way with the aid of train location, which might be provided by GPS, accelerometer, goniometer and so on[3].

However, positioning techniques applied to wireless communication generally occupy communication resources, whether it's CSI method, DoA and ToA estimation, beam sweeping or satellite communication. However, mobile communication users like pedestrians and cars could be detected and positioned by computer vision especially in Line of Sight (LOS) scenarios. So computer vision might be a practical way to provide location-awareness support. In [4], an image tracking algorithm of binocular vision in LOS scenario is put forward to track users for beamforming. A position-based BF system employing 3D image reconstruction with the assistant of co-located cameras has been proposed as well and it has been proved that acquiring location information by cameras can achieve higher spectrum efficiency in beamforming systems[5]. [4] and [5] were our previous work, which carried out simple geometric models but did not employ CV to detect object location.

Object detection is one of the important problems in computer vision, which is aimed to achieve tasks of object classification and localization in images or videos, the former of which may indicate potential mobile users like pedestrians or cars, the letter of which can indicate the position of users. And convolutional neural network (CNN) takes a place in this field because of its ability to extract two-dimensional correlation features through 2D conventional operation. Regions with CNN features (R-CNN) can pick out regions that might include interested objects and process them with CNN[6]. You-only-look-once (YOLO) network divides images into several cells and process the whole image with CNN to classify and localize objects in each cell[7]. In this paper, YOLOv3 is used for object detection, which performs well especially in detecting small objects[8].

In this paper, we introduce a location-aware beamforming scheme based on computer vision in outdoor LOS scenarios. With much lower or even without any wireless overhead for positioning or beam measurement/tracking, cameras co-located with antenna

Sponsored by National Science and Technology Major Project with grant No. 2018ZX03001024-006.

arrays combined with CNN will estimate mobile terminals' location information precisely and quickly. Different from those typical CV approaches, the proposed method utilizes intentionally blurred or low-resolution video/images for faster processing while preserving certain level of user privacy. The method would promise a great improvement in terms of time delay and spectrum efficiency.

The remainder of the paper is organized as follows. Section II introduces the system model we built for computer-vision based beamforming. Section III presents the simulation results of our model and analyses the antenna array and beam configuration as well as the location error of the system. Then future work is listed in section IV, including CNN trained for communication purpose, high mobility support and more information extraction with CNN. At last, section V makes a conclusion about the this paper's work.

## II. System Model

### A. Beamforming Coverage Scenarios

We consider the base station has an antenna array of height $h$ covering a LOS region for radius $r$ from $r_1$ to $r_2$ and azimuth angle $\varphi$ from $\varphi_1$ to $\varphi_2$. $M$ x $M$ antennas are used to get space dimensional gain, which is, in this paper, array gain from analog beamforming. The camera that captures videos to aid the BS perform location-based beamforming is assumed to be co-located with BS antenna array. Only pedestrians' mobility is considered in this paper, and we assume that they move at a low speed, e.g. 3km/h.

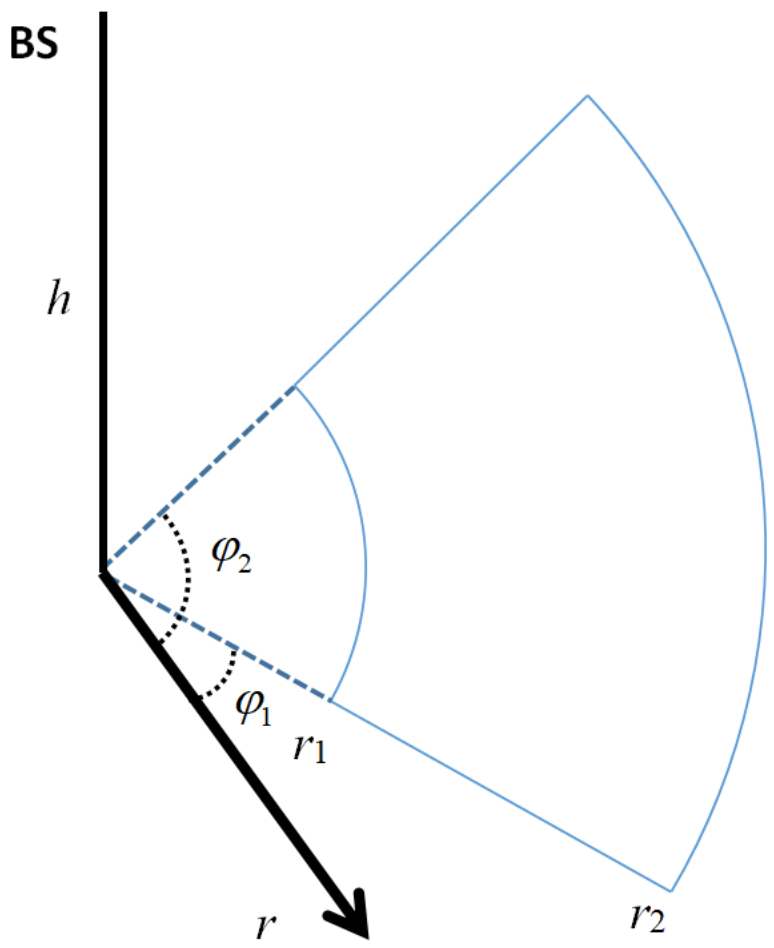

Figure 1. Coverage scenario

### B. Channel Model

*1) Free-Space Path Loss:* Since LOS scenarios with few arriers are considered in this paper, the effect of scattered multipath is not taken into account. So we apply free-space path loss model which can be expressed as:

$$PL = -32.4 - 20\log(f) - 20\log\left(\frac{d}{1000}\right) \quad (1)$$

where $f$ represents frequency of the signal, measured by MHz and here it's 28GHz, a popular frequency of millimeter wave. $d$ is the 3D distance between BS antenna array and the mobile users, measured by meters, and it's expressed as:

$$d = \sqrt{h^2 + r^2} \quad (2)$$

*2) Antenna Gain:* According to [9], antenna gain is expressed as:

$$G_a = G_m - 12\times\left(\frac{\varphi'}{\varphi_{3\mathrm{dB}}}\right)^2 - 12\times\left(\frac{\theta'}{\theta_{3\mathrm{dB}}}\right)^2 \quad (3)$$

where $\varphi'$ and $\theta'$ represent azimuth and zenith angle of user in the coordinate system of antenna array, $\varphi_{3\mathrm{dB}}$ and $\theta_{3\mathrm{dB}}$ represent horizontal and vertical half power angle of antenna and Gm is the maximum directivity gain. Since we aim to ensure coverage performance of a LOS region instead of typical wide areas, antennas are pointed to the middle of the region.

*3) Beamforming Scheme:* When coverage scenario determined, a total number of $N$ beams are generated, each of which points to different locations. Due to LOS scenarios, each beam can be considered to be an optimal solution.

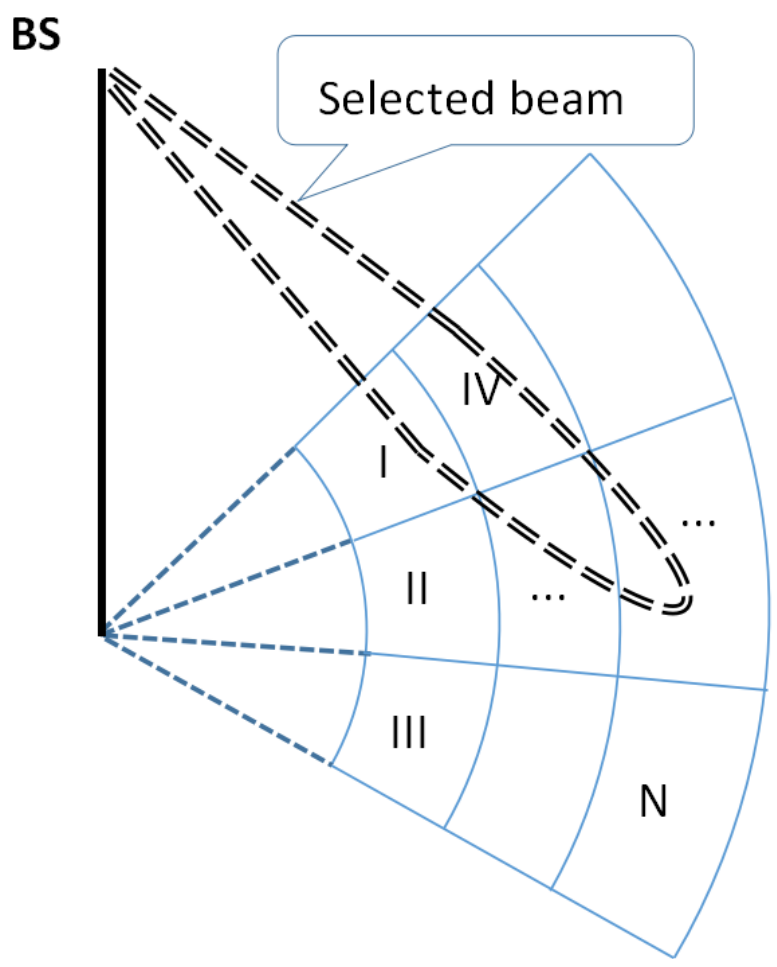

Figure 2. Coverage with one of $N$ beams

When the system starts to work, camera captures real-time videos constantly as inputs to CNN. Then CNN carries out the detection and localization of certain objects, such as the pedestrians communicating with the BS. And beams are selected according to the location of objects. The process above is operated repeatedly as long as the mobile terminal carried by the pedestrian has an active radio link with the BS. It should be noted that initial access process and the alignment of wireless beam with certain object detected from the videos are not considered but will be studied in our future work.

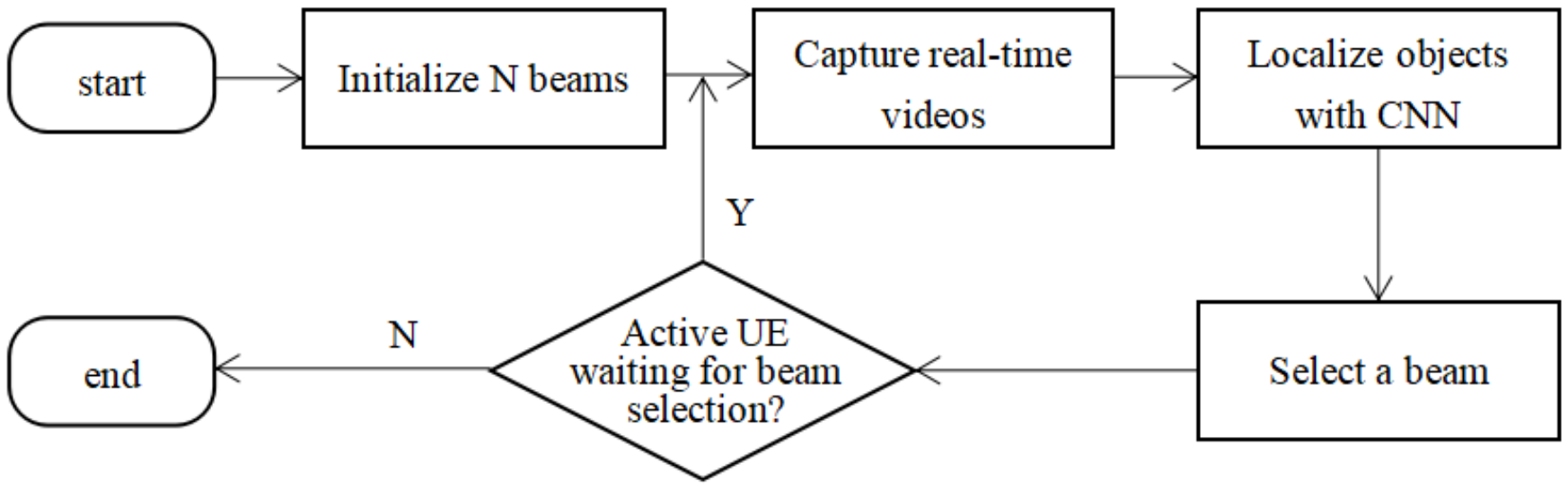

Figure 2. Beam selection process

For quantitive analysis, we utilize the beamforming model in [9] to obtain beamforming gain GBF.

### C. *Location Error Analysis*

The total location error indicator of CNN-based localization system could be defined as:

$$E = E_1 + E_2 \tag{4}$$

The indicator $E$ is defined to describe how accurate a visual localization could be for beam selection advice and it has two components. $E_1$ represents the location deviation caused by movement of mobile terminal, and $E_2$ represents the positioning error of CNN. In order to evaluate the positioning error of a CNN, we only concentrate on the coordinate of an object instead of bounding box in typical CV methods. CNN gives detection results of each mobile terminal at each frame, however, with processing delay. Therefore, mobile terminals may move to a different location after a period of delay, leading to location deviation mentioned before. And though a mobile terminal is absolutely still, CNN is not able to detect its accurate position without any error. We also assume that if a certain kind of CNN is determined, the processing delay is fixed to a certain value, and thus the former component only depends on the moving speed of a mobile terminal. So these two parts of the total error are considered to be independent of each other under a given configuration in our model.

Moreover, the total error can also be decomposed in two one-dimensional errors, one in radial direction and one in tangential direction, expressed as below respectively:

$$E_r = \frac{v_r}{F} + K_h E_{2h} \tag{5}$$

$$E_t = \frac{v_t}{F} + K_w E_{2w} \tag{6}$$

where $E_{2h}$ and $E_{2w}$ represent location error in height and width direction performed by CNN, measured by pixels. Since the beamforming selection process mainly depends on the coordinate of the user, we only take the offset of bounding box as location error, different to typical error evaluation of detection task. $K_h$ and $K_w$ achieve a conversion from pixel to actual length, which may differ with the location of camera, shooting angles and resolution. $F$ represents the frame rates of CNN output in frame per second, which is assumed to match the input video frame rates.

Moreover, it is noticed that there is a trade-off between two parts of the error by varying the system configurations. In general, the processing speed and accuracy is a couple of contradiction for most kinds of CNN, each contributing different parts of the location error in our system. So there might be a better CNN to achieve lower total error for some certain scenario. Since we only use one type of CNN, this issue remains for further study.

Besides, videos could be blurred to protect users' privacy with lower positioning accuracy. If CNN and coverage scenario determined, a best blur level can be chosen to fit a certain processing speed with the aim of both meeting accuracy of beamforming and protecting privacy.

### D. *Evaluation Method*

Assuming transmit power is $P_0$, the received signal strength can be expressed as:

$$P = P_0 + G_a + G_{BF} + PL \tag{7}$$

This would be the reference indicator of coverage performance in this paper.

Ideal coverage performance will be discussed first, where we assume that mobile terminals are stationary and visual localization system has no error. We'll investigate the antenna and beam configuration with the aim to meet different signal strength requirements. Then we'll obtain location error for videos of different Gaussian blur levels, in order to find the best blur level for both protecting users' privacy and meeting accuracy requirements. Finally, we'll observe the influence of location error obtained previously, and discuss the effect of mobility and CNN accuracy for some traces of nodes with certain mobility pattern. [5] has proved the improvement of spectrum efficiency in beamforming systems with the help of cameras to get terminals' location information, which will not be discussed in this paper. And we will evaluate the time delay, in other

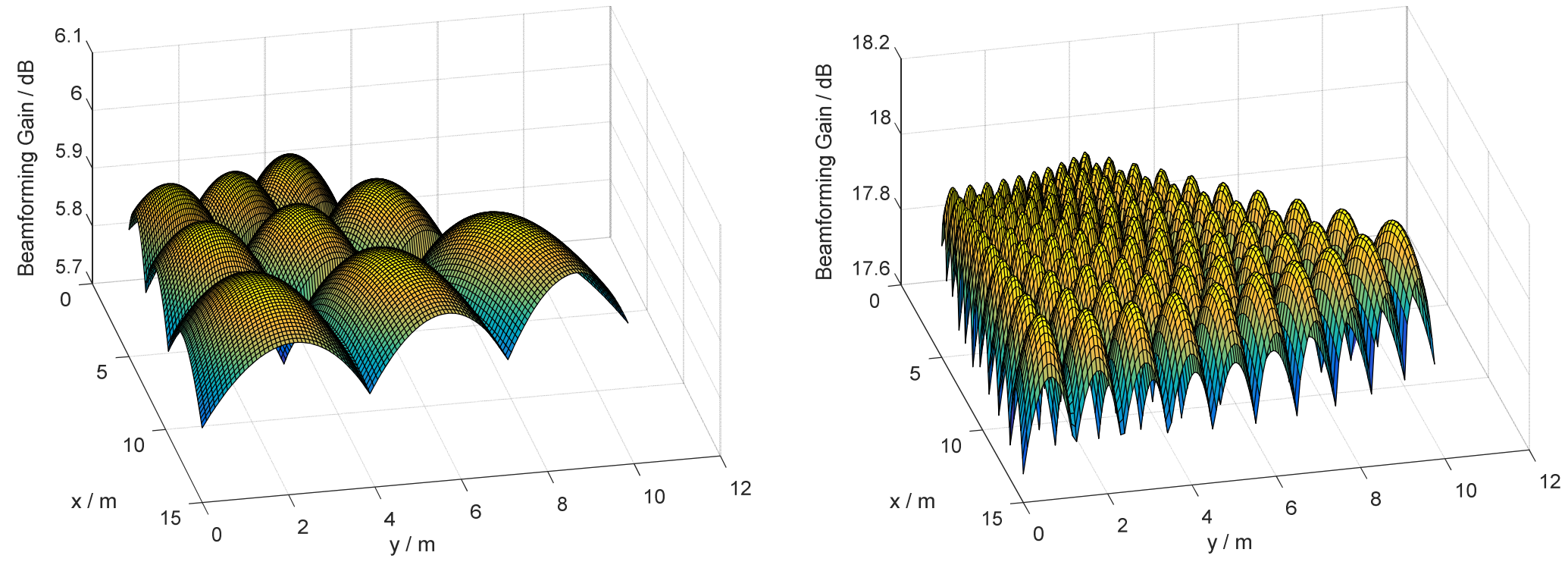

(a) 2x2 antennas with 9 beams (b) 8x8 antennas with 100 beams

Figure 4. Beamforming gain without location error

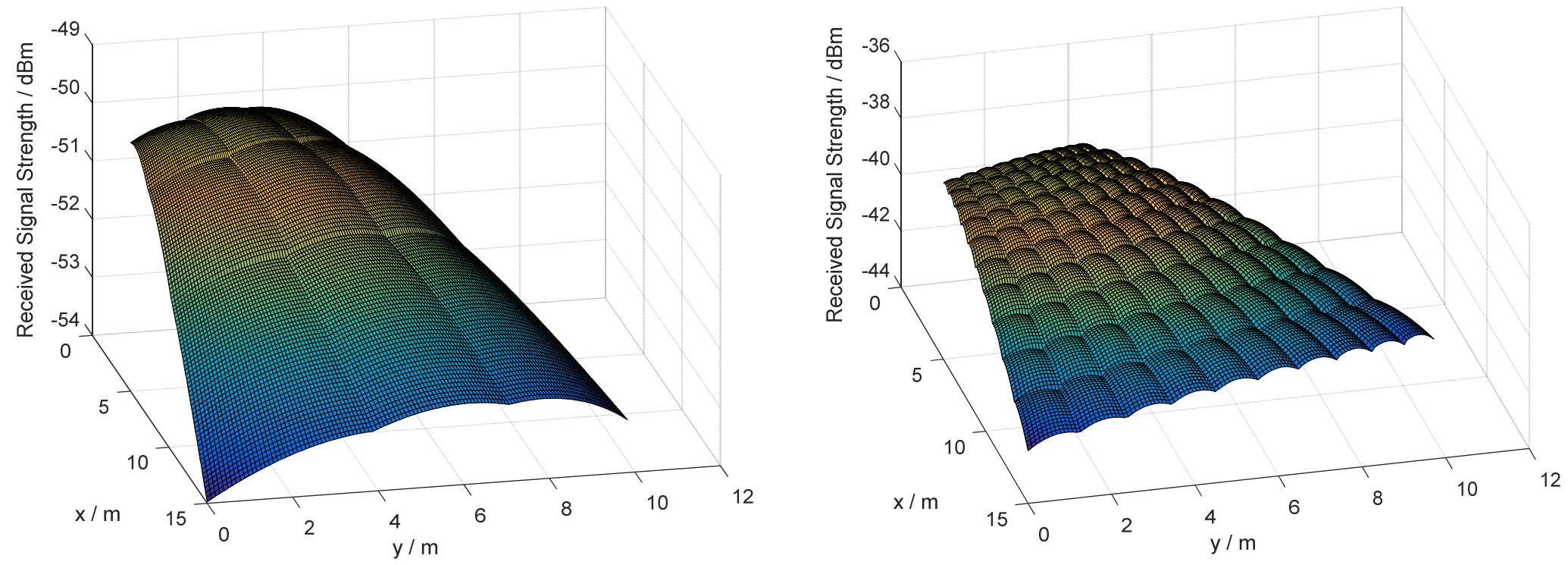

(a) 2x2 antennas with 9 beams (b) 8x8 antennas with 100 beams

Figure 5. Received signal strength without location error

words, beam tracking speed by observing the signal coverage strength under different mobility.

## III. Simulation Results and Discussions

Simulation parameters are listed in table I.

TABLE I. Simulation Parameters

| $h$ | $r$ | $\phi$ | $P_0$ | $\theta_{3dB}$ |
|---|---|---|---|---|
| 10m | 5-15m | 0-45degree | 20dBm | 65degree[a] |
| $\varphi_{3dB}$ | $G_m$ | $f$ | $F$ | $v$ |
| 65degree[b] | 8dB[c] | 28GHz | 10FPS | 0-1.5m/s |

a.b.c. recommended by [9]

### A. Ideal Coverage Performance

First we exhibit the simulation results of coverage performance without location error. Figure 4. shows the beamforming gain with 2x2 antennas and 8x8 antennas. The BS is at the origin with height of $h$. It is noticed that the number of beams is determined by the number of antennas with the aim of approaching expected beamforming gain. For example, with 2x2 antennas, 9 beams are enough to ensure beamforming gain above 5.8dB, only 0.2dB lower than 6dB; for 8x8 antennas, 100 beams are needed, which is able to maintain beamforming gain above 17.6dB, 0.4dB lower than 18dB. When more antennas are used to achieve higher beamforming gain, beams become narrower, and therefore, more beams are required for a more balanced coverage.

Then corresponding coverage in terms of received signal strength is presented by figure 5.

### B. Location Error

The same video from [10] is intentionally blurred by different levels of Gaussian blur as inputs to the YOLOv3 network. Gaussian blur is operated through a 2-D convolutional kernel, which is generated by two dimensional Gaussian function[11]. And a larger standard deviation of the Gaussian function results in larger blurring level, also promising greater protection to users' privacy as presented in figure 6.

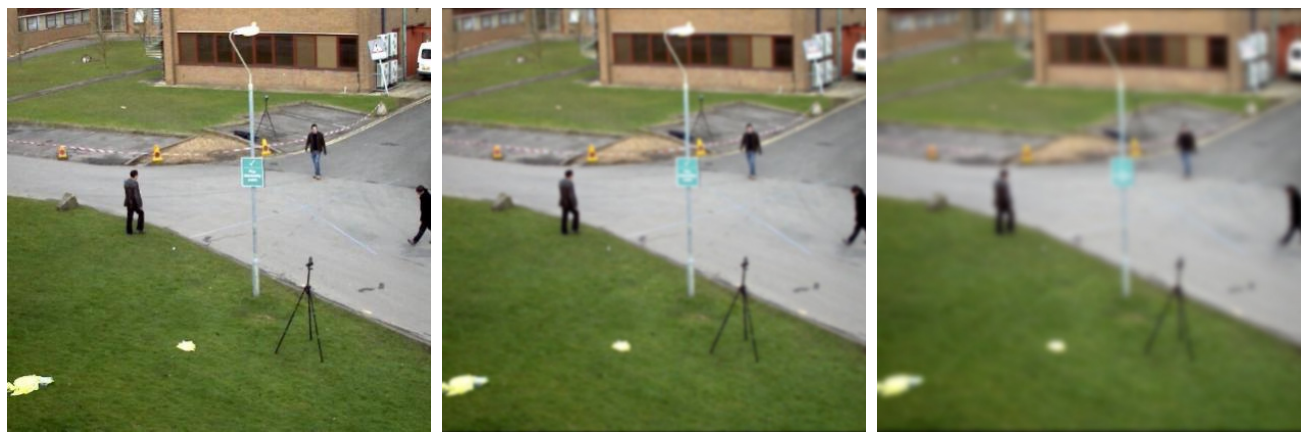

(a) Original (b) Blur level 5 (c) Blur level 10

Figure 6. Videos blurred by different levels of Gaussian blur

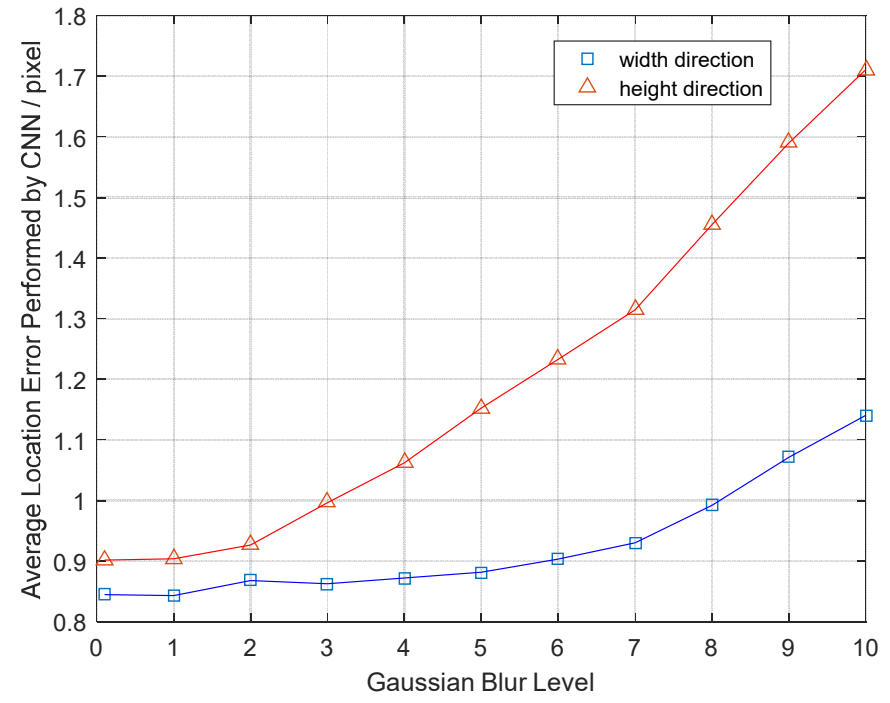

Figure 7. Location error performed by CNN

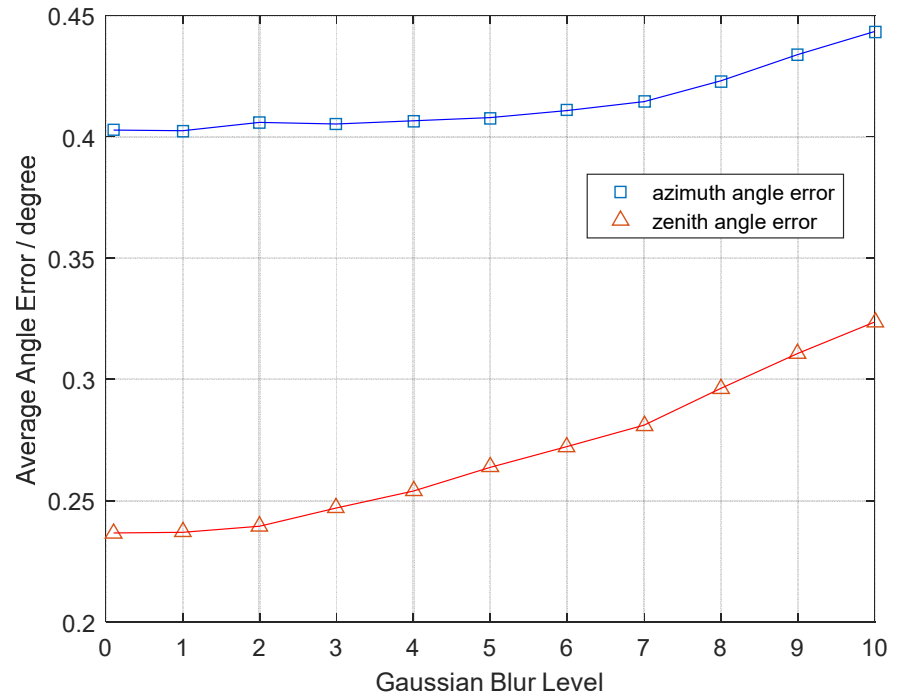

Figure 8. Average angle error

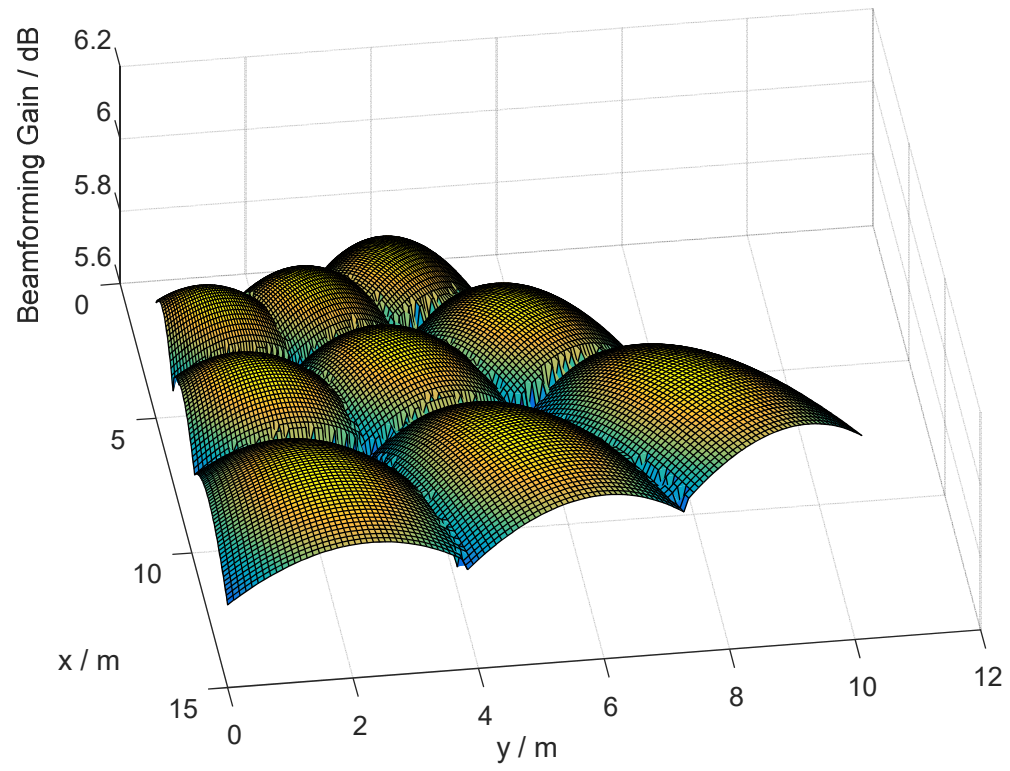

Figure 9. Beamforming gain with location error

Then we get ground location errors along radial and tangential directions for different Gaussian blur levels through YOLOv3 network, as presented in figure 7. Assuming radius to be 10m and maximum velocity to be 0.5m/s, the average azimuth and zenith angle errors are shown in figure 8.

If maximum angle error given, the highest blur level can be set, in order to not only protect users' privacy, but also ensure the accuracy requirements for beamforming. For example, if a beamforming system requires azimuth angle error below 0.42 degree and zenith angle error below 0.3 degree, a Gaussian blur level below level 7 will meet the

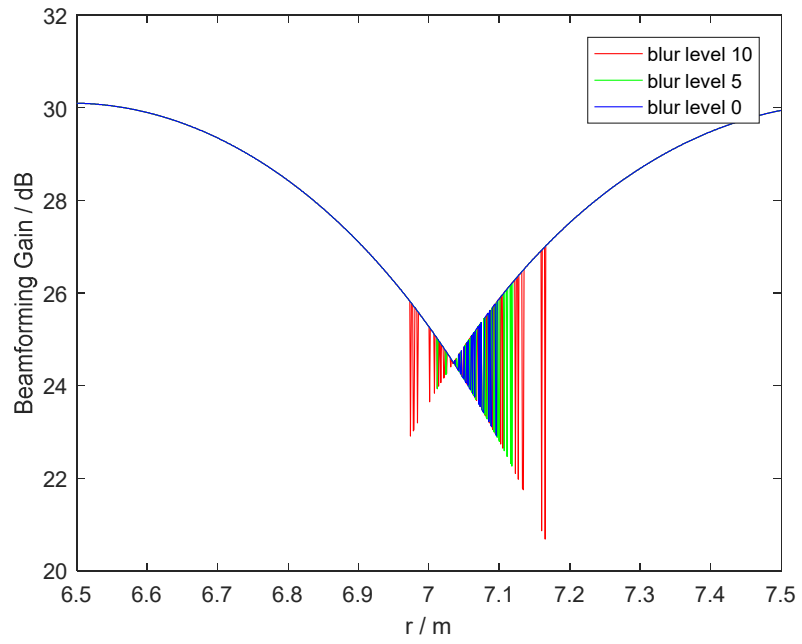

(a) radial direction, $\varphi = \dfrac{\pi}{8}$

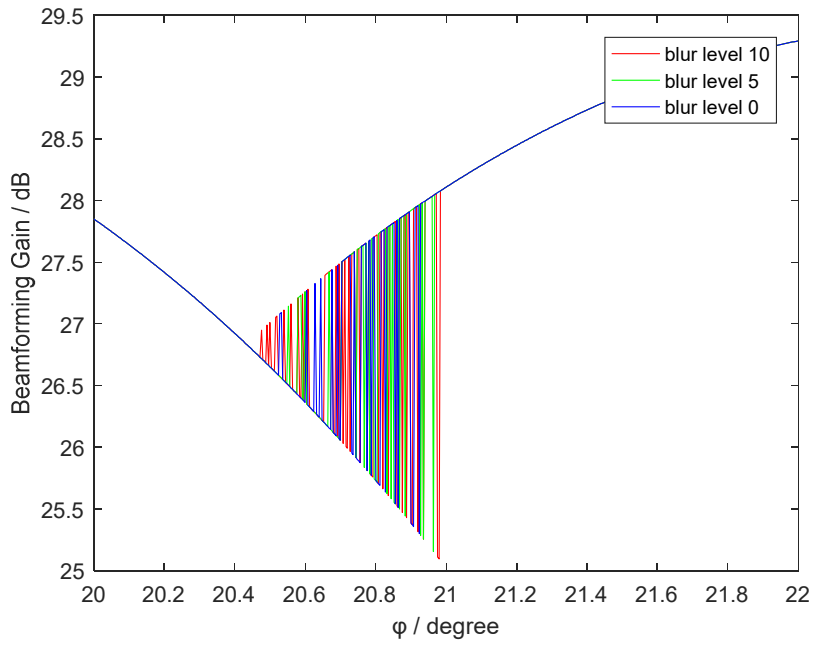

(b) tangential direction, $r$=10m

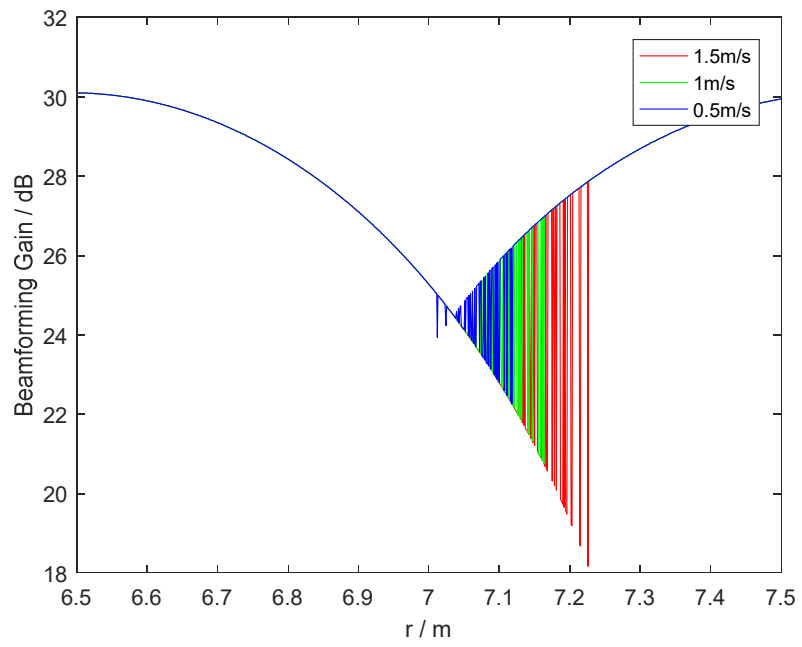

(c) radial direction, $\varphi = \dfrac{\pi}{8}$

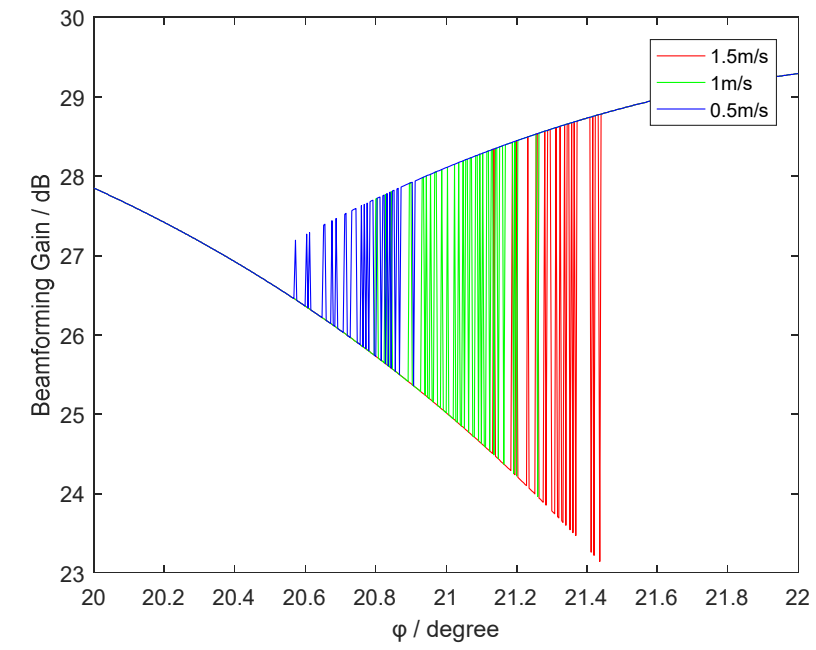

(d) tangential direction, $r$=10m

Figure 10. Beamforming gain with different speed and blur levels in two paths

requirements.

### C. Coverage Performance with Location Error

We then exhibit the beamforming gain with the location error as in Figure 9, where the antenna configuration is 2×2 and 9 beams are generated. It is observed that at the edges of cell coverage regions, there are decreases in gain, which is caused by misalignment of beams, therefore less optimal beams are selected.

In order to observe how the proposed scheme performs at the edges of regions as well as how two components of the total error affect coverage performance at the edges, we set antenna and beam configuration to a relatively extreme situation, 32×32 antennas with 77 (11×7) beams, where signal strength falls quickly near the edges, making the effect of misalignment more apparent. We also assume that mobile users walk through the region at a certain speed by two paths, one of which is along radial direction, another of which is along tangential direction. Figure 10. (a) and (b) show the influence of different location error obtained by the CNN; (c) and (d) present the influence of different walking speed.

It is more obvious that at the edge of regions, misalignment results in signal strength decreases. Moreover, as presented in figure 10, as moving speed gets higher or CNN outputs get less accurate, misalignment may appear in a larger range, which causes worse performance at the edge. So there is a trade-off between moving speed and accuracy. In this case, for example, in radial direction, if the beamforming gain requirement is above 22dB and speed is no higher than 0.5m/s, then videos of Gaussian blur level 5 or lower are required. And in turn, if the camera can only get videos no more clearer than Gaussian blur level 5, the system won't be able to support scenarios that users move faster than 0.5m/s.

And at blur level 5, the signal strength decreases by no more than 4dB at 0.5m/s, 6dB at 1m/s and 10dB at 1.5m/s. Even the configuration is relatively extreme, our system still performs well under low mobility in terms of time delay.

Our work in this paper is still preliminary, and there are many interesting and important issues that need to be addressed before the proposed scheme could be practically implemented, if possible. And these issues are briefly described below.

## IV. Future Work

### A. CNN Trained for Communication Purpose

In our current model, beam selection relies on the detection results of CNN, which is trained by typical computer vision methods. In other words, existing CNN detection systems tend to be consistent with human visual perception rather than cater to communication purposes.We assume that if a visual detection system were trained by some wireless indicators such as channel quality, it would be able to provide better recommendations for beamforming. And this type of CNN would still achieve detection, not working like human eyes anymore but eyes for wireless communication like radars. Going further, this CNN would be different from typical CNN detectors in the following aspects.

*1) Network output is mainly used for communication purpose.* In typical CNN detectors, network output includes whether there is an object or not, object classes and bounding boxes indicating the coordinate and size of an object. It is noticed that there is a huge waste in network output, which leads to deeper network and lower efficiency. We suppose a communication-aiding CNN should be able to output instructions for wireless communication such as beam selection or wireless channel scenarios, or detect whether the objects is related to wireless communication or not to avoid unnecessary radiation or interference. And for the case we consider in this paper, network output could be beam number. For the new output, the neural network structure may also need to be redesigned.

*2) Loss function redefinition.*Since the neural network output is redesigned, loss function has to be redefined. Typical detection methods concentrate on the offset and size error of bounding box as well as prediction error of classes. In order to train the CNN we propose, for example, a CNN that gives beam selection recommendation for a single user, where generated beams each coverage a region that is numbered by rows and columns, the loss function could be simply defined as:

$$L = \lambda\left[\left(m - \hat{m}\right)^2 + \left(n - \hat{n}\right)^2\right] \tag{8}$$

where $m$ and $n$ represent the row number and column number of the optimal beam, and $\hat{m}$ and $\hat{n}$ represent row number and column number of the predicted beam. Similar to actual distance, the loss function describes distance between the optimal beam and the predicted beam. If optimized successfully, the predicted beam is expected to be close to the optimal one.

*3) Real-time training.*

Training labels should be obtained by wireless channel in real time instead of prior manual labeling. For instance, the optimal beam selection could be determined by signal strength feedback of mobile terminals and be used immediately in training. And this would be a CNN closely coupled with wireless communication.

### B. High Mobility Support

Mobile terminals with high mobility are not considered in this paper because we find that the CNN error significantly increases when mobile users move at a high speed. There comes two ways to reduce the impact of high mobility. First, an efficient way is just to improve the processing speed of CNN. Variety types of CNN might be tested to adapt to high mobility scenarios. Due to the trade-off relationship mentioned in section II, between processing speed and localization precision, which are determined by the type of CNN, are supposed to achieve a best balance to minimize the total error. Further, our future communication-aiding CNN will also be tested and we expect it to be suitable for high mobility scenario because of its lightweight and non-wasteful network structure.

The second way is to introduce tracking algorithm to CNN detectors. Based on YOLO real-time detection, a tracking algorithm has been proposed in [12], where researchers obtain and predict temporal relationship between frames with long short-term memory (LSTM). And we suppose that the introduction of tracking algorithms might reduce unexpected mobile user deviation between frames. It should be noted that the channel state information extracted from wireless communication might be useful for supporting better object-tracking ability of traditional CNN, just like the CNN designed for processing visible light images/videos could be more powerful with the help of a radar.

### C. *Extracting More Information*

More information about users could be extracted, such as moving patterns in different areas, congestion of pedestrians and vehicles and hot spots that mobile users prefer to stay. All these information could also be obtained by multiple cameras, as well as shared and co-calculated by the way of cloud computing or edge computing. The computer vision system owned by wireless operators would be able to give advice for scheduling or even network planning in mobile communication, such as to optimize the synchronous or control channels' coverage dynamically according to the users distribution.

## V. CONCLUSIONS

In this paper, we put forward a location-aware beamforming scheme with the help of computer vision. The location information is provided by detection results of CNN, rather than wireless overhead or GPS. That is expected to save wireless communication resources, which can increase spectrum efficiency and decrease time delay.

A model has been built to investigate the coverage performance, and we have obtained location error in a certain scenario by using CNN. Simulations indicate that the beamforming scheme is practical and the CNN we use is sufficient in both terms of accuracy and processing speed required by beamforming. And potential future work is discussed.